\documentclass[aps,showpacs,amsmath,amssymb,twocolumn,final]{revtex4-1}

\usepackage{graphicx, epstopdf}

\usepackage{dcolumn}% Align table columns on decimal point
\usepackage{bm}% bold math
\usepackage{ulem}
\usepackage{soul}

\usepackage{dcolumn}% Align table columns on decimal point
\usepackage{bm}% bold math

\begin{document}

%\preprint{}

\title{Nano-bridge Superconducting Quantum Interference Devices: beyond the Josephson limit }

\author{Dibyendu Hazra}
\affiliation{Department of Physics, Indian Institute of Technology Kanpur, Kanpur-208016, India }
\date{\today}

\begin{abstract}
Nano-scale superconducting quantum interference devices (nano-SQUIDS) where the weak-links are made from nano-bridges --- i.e., nano-bridge--SQUIDs (NBSs) --- are one of the most sensitive magnetometers for nano-scale magnetometry. Because of very strong non-linearity in the nano-bridge--electrode joints, the applied magnetic flux ($\Phi_{a}$) -- critical current ($I_{c}$) characteristics of NBSs differ very significantly from conventional tunnel-junction-SQUIDs, especially when nano-bridges are long and/or the screening parameter is large. However, in most of the theoretical descriptions, NBSs have been treated like conventional tunnel-junction-SQUIDs, which are based on d.c. Josephson effect. Here, I present a model demonstrating that for long nano-bridges and/or large screening parameter the $I_{c}(\Phi_{a})$ of a NBS can be explained by merely considering the fluxoid quantization in the NBS loop and the energy of the NBS; it is not necessary to take the Josephson effect into consideration. I also demonstrate that using the model, we can derive useful expressions like modulation depth and transfer function. I also discuss the role of kinetic inductance fraction ($\kappa$) in determining $I_{c}(\Phi_{a})$.        

\end{abstract}

%\pacs{74.78.-w, 73.63.-b,74.81.Bd}
%\keywords{Superconducting weak links, Hysteresis, SQUID}

\maketitle

\section{Introduction}

Nano-SQUIDs are the most sensitive magnetometers to measure the magnetic properties of individual nano-particles or to probe the local magnetic properties of a sample in the sub-micron scale \cite{wernsdorfer2007classical,wernsdorfer2009micro,foley2009nanosquids,vasyukov2013scanning, levenson2013dispersive, granata2016nano, gallop2016nanoscale,yue2017sensitive}. The other applications of nano-SQUIDs include measuring persistent current in a phase coherent ring \cite{mailly1993experimental, rabaud2001persistent}, single-photon detection \cite{hao2003inductive}, detecting motion of a nano-mechanical oscillator \cite{etaki2008motion} and as  non-linear circuit-elements in quantum bits \cite{vijay2009optimizing}. Consequently, nano-SQUIDs have been developed from versatile methods and by using different types of weak-links (WLs) \cite{granata2016nano, martinez2016nanosquids}, like, nano-bridges (NBs) \cite{hasselbach2002micro, lam2003development, troeman2007nanosquids, hao2008measurement, vijay2010approaching, mandal2011diamond,hazra2013nano,chen2016high}, superconductor--normal-metal--superconductor (SNS) proximity junctions \cite{angers2008proximity, ronzani2013micro, samaddar2013niobium}, tunnel junctions (TJs) \cite{wolbing2013nb, granata2013three, schmelz2016nearly}, and carbon nano-tube \cite{cleuziou2006carbon} to mention only a few. Out of these, NBSs have been most commonly used primarily because of their easy fabrication method \cite{wernsdorfer2009micro, granata2016nano}.         

Conventionally, a d.c. SQUID operation has been understood based on two phenomena: The d.c. Josephson effect and the fluxoid quantization in a superconducting loop \cite{clarke2006squid}. An ideal d.c. Josephson effect predicts the flow of a loss-less current --- the supercurrent, $I_{s}$  --- between two superconductors interrupted by a WL. $I_{s}$ follows the relation: $I_{s} =I_{c}sin(\theta)$, where $I_c$ is the critical current and $\theta$ is the phase of the WL. This relation holds provided most of the phase across the superconductor--WL--superconductor drops between the WL, resulting in a well-defined phase of the WL, for instance, as it happens in TJs \cite{likharev1979superconducting,tinkham1996introduction}. In case of a NB, the phase of the bridge is not well-defined in most of the cases  \cite{likharev1979superconducting,hasselbach2002micro,podd2007micro,gumann2007microscopic}. The ideal Josephson relation in NBs, therefore, only manifests in limiting cases, e.g., where bridge dimensions are smaller than the temperature dependent Ginzburg-Landau coherence length ($\xi_{T}$) \cite{likharev1979superconducting,tinkham1996introduction,hasselbach2002micro}.   

Consequently, in NBSs, various features in the $I_{c}$($\Phi_{a}$) have been observed --- for instances, triangualar-shaped \cite{hasselbach2000microsquid,faucher2002niobium, hasselbach2002micro,hutchinson2003hot,hutchinson2004controlled,hutchinson2004fabrication, mandal2011diamond,hazra2013nano,russo2012nanoparticle, hazra2014nano,paul2016micron,mccaughan2016nanosquid, wu2017measurement, biswas2017josephson}, double-branched  \cite{hasselbach2000microsquid,faucher2002niobium, hasselbach2002micro,hutchinson2003hot,hutchinson2004controlled,hutchinson2004fabrication, hazra2014nano,russo2016nanosquids} and  a diamond-shaped $I_{c}$ ($\Phi_{a}$) \cite{faucher2002niobium, hasselbach2002micro,hazra2013nano,russo2016nanosquids} ---   which are not conceivable by a conventional d.c. SQUID theory \cite{clarke2006squid,granata2016nano}. Thus, alternative theories \cite{hasselbach2002micro,podd2007micro} have been developed which describe some of the features, like, the non-sinusoidal $I_{c}$($\Phi_{a}$). 

Here, I present a model that explains all of the above mentioned experimental features. More importantly, unlike the previous models, here, I demonstrate that for a NBS with long nano-bridges and/or large screening parameter, the fluxoid quantization in the NBS loop and the energy of the NBS can explain all the experimental features of $I_{c}$($\Phi_{a}$), without considering the Josephson effect. Moreover, the model presented here derives the expression for modulation depth and transfer function.

\section{Model of a nano-bridge--SQUID beyond the Josephson limit}

\begin{figure}\centerline{\includegraphics[width=9cm,angle=0]{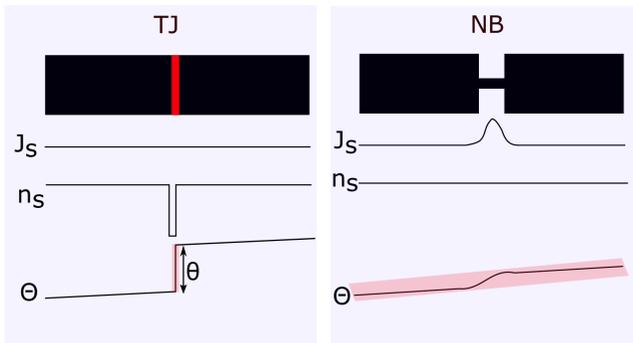}}
	\caption {Schematics showing the spatial distribution of supercurrent density, $j_s$, Cooper pair density, $n_s$ and phase, $\Theta$, for a tunnel junction (TJ) and for a nano-bridge (NB). In case of a TJ, there is a sharp drop of $\Theta$ across the junction, making $\theta$ well-defined. In case of a NB, $\Theta$ spreads almost uniformly across the whole structure, resulting in a poorly defined $\theta$. }
	\label{fig:phase}
\end{figure}

I start by presenting a qualitative comparison between a TJ and a NB--- how the phase ($\Theta$) of the superconducting order parameter is distributed in these two cases, in presence of a finite $I_s$. In presence of a finite $I_s$, $\Theta$ is spatially non-uniform and the phase gradient is related to the supercurrent density ($j_s$) and the Cooper-pair density ($n_s$): $\nabla \Theta \propto j_{s}/n_{s}$ \cite{tinkham1996introduction}. In a TJ, the insulating layer has negligible Cooper-pair density: $n_{s} \to 0$, the most of the $\Theta$ drops across the insulating layer, yielding a well-defined $\theta$, as shown in Fig.\ref{fig:phase}. In case of a NB, the bridge and the electrodes being made of the same superconductors, $n_{s}$ is almost same in NBs and in electrodes. The enhancement of the phase gradient in the NB is the result of the enhancement of $j_{s}$ due to the smallness of the width of the NB in comparison to the adjacent electrodes. In practical NBSs, the width of the NB is made typically $\sim $ 2--3  times smaller than the adjacent electrodes (much wider electrodes are not desirable in order to avoid vortex penetration). Moreover, for a long NB, i.e., when the NB is longer than $\xi_{T}$, $j_{s}$ increases smoothly near NB--electrode joint \cite{likharev1979superconducting}. Altogether, in a typical NB, unlike a TJ, the phase-drop across the NB is of the same order as the phase-drop in the electrodes, resulting a poorly defined $\theta$. This can also be viewed as if $\theta$ is spread beyond the NB deep inside the electrodes \cite{likharev1979superconducting,hasselbach2002micro,podd2007micro,gumann2007microscopic}, allowing to treat a NB just like its electrodes with a smaller critical current. In Fig.\ref{fig:phase}, I juxtapose a NB alongside a TJ in order to compare the spatial variation of $j_s$, $n_s$ and $\Theta$ in these two types of WLs.     

\begin{figure}\centerline{\includegraphics[width=6cm,angle=0]{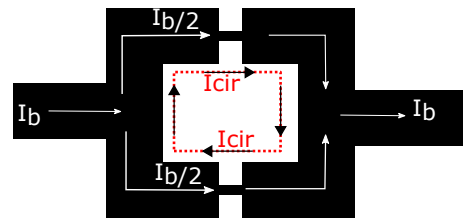}}
	\caption {Schematic of a symmetric NBS. The bias current, $I_b$, and the circulating current, $I_{cir}$, are shown by arrows. Here, $I_{cir}$ is shown clockwise, but depending on fluxoid number, $n$, it may also circulate counterclockwise. }
	\label{fig:nanoSQD}
\end{figure}

Now, let me consider a standard NBS geometry, as shown in Fig.\ref{fig:nanoSQD}. Here, I consider a symmetric NBS where both NBs have the identical critical current, $I_c$; the asymmetric case can be straightforwardly generalized. When the NBS is biased with a dc current, $I_b$, it splits equally into two parallel branches--- a current $I_b/2$ flows across each NB. That apart, due to the fluxoid quantization in the NBS loop, another current,  $I_{cir}$, may circulate, especially, in presence of a finite $\Phi_{a}$. In Fig.\ref{fig:nanoSQD}, I have schematically shown both $I_b$ and $I_{cir}$. Clearly, $I_{cir}$ breaks the symmetry of the net current flow in two branches--- now, the net current flowing across two NBs are $\frac{I_{b}}{2} + I_{cir}$ and $\frac{I_{b}}{2} - I_{cir}$, respectively. Starting from zero, with increasing $I_b$, depending on $I_{cir}$, the net current flow across one or both the NBs will be $I_c$, at a particular bias current. I assume that if the net current flow across, at least, one of the NBs become $I_c$, it immediately switches to the voltage state--- the corresponding $I_{b}$ is identified as the critical current, $I_{cs}$, of the NBS. Therefore, $\frac{I_{cs}}{2} + |I_{cir}| = I_{c} $. Rearranging, $I_{cs}$ can be written as 

\begin{eqnarray}
I_{cs} = 2(I_{c} - |I_{cir}|).
\label{eq:Ics}
\end{eqnarray}

Note that, maximum $I_{cs}$ is $2I_{c}$, i.e., when  $I_{cir} = 0$ and the net current flow across both the NBs becomes $I_{c}$.   
 
For a given $\Phi_{a}$, $I_{cir}$ can be evaluated from the fluxoid quantization formula       

\begin{eqnarray}
L{_t}I_{cir} +  \Phi_{a} = n\Phi_{0}, 
\label{eq:FQ}
\end{eqnarray}

here,  $L{_t} = L{_l} + L{_k}$, is the total inductance of the NBS; $L{_l}$ and $L{_k}$ are loop and kinetic inductance, respectively. The origin of the $L{_k}$ is the kinetic energy due to the motion of the Cooper pairs \cite{henkels1977penetration,barone1982physics}. $n$ is an integer and $\Phi_{0}$ is the flux quanta. The magnitude and sign (sense of circulation) of $I_{cir}$ depend on $n$.  

For a given $\Phi_{a}$, $n$ can have multiple values--- the most probable $n$ corresponds to the minimum energy ($E$) of the NBS which can be written as

\begin{eqnarray}
\nonumber E = \frac{1}{2}L_{k} \left[ \left( \frac{I_{b}}{2}+I_{cir}\right)^{2}+ \left(\frac{I_{b}}{2}-I_{cir}\right)^{2}  \right] + \\*
      \frac{1}{2}L_{l}I_{cir}^{2}. 
\label{eq:E}
\end{eqnarray}

The first term within square bracket is the kinetic energy of the Cooper pairs; the second term is the magnetic energy due to the circulation current. Moreover, to remain in the superconducting (zero-voltage) state, $|I_{cir}|$ cannot exceed $I_{c}$. This imposes  restrictions on $n$, following Eq.\ref{eq:FQ}: 

\begin{eqnarray}
|\frac{n\Phi_{0} - \Phi_{a}}{L{_t}}| \le I_{c}.  
\label{eq:n}
\end{eqnarray}  

Eqs.\ref{eq:Ics}--\ref{eq:n} lay the foundation to understand $I_{c} (\Phi_{a})$ of NBSs beyond the Josephson limit. It is convinient to express Eqs.\ref{eq:Ics}--\ref{eq:n} in terms of dimensionless units. I normalize currents by the maximum critical current of the NBS $I_{0} = 2I_{c}$, magnetic flux by $\Phi_{0}$, and the energy by $\frac{1}{2}L_{k}I_{0}^{2}$. With these normalizations, Eqs.\ref{eq:Ics}--\ref{eq:n} take the form:    

\begin{eqnarray}
i_{cs} = (1 - 2|i_{cir}|),
\label{eq:icss}
\end{eqnarray}

\begin{eqnarray}
i_{cir} = \frac{n - \phi_{a}}{\beta_{L}} ,  
\label{eq:fqq}
\end{eqnarray} 

\begin{eqnarray}
\nonumber \epsilon  =  \left[ \left( \frac{i_{b}}{2}+i_{cir}\right)^{2}+ \left(\frac{i_{b}}{2}-i_{cir}\right)^{2}  \right] + \\*
\frac{(1-\kappa)}{\kappa} i_{cir}^{2},  
\label{eq:ee}
\end{eqnarray}

and 

\begin{eqnarray}
|n- \phi_{a}| \le \frac{\beta_{L}}{2},  
\label{eq:nn}
\end{eqnarray} 
respectively.

Here, $i_{cs} = I_{cs}/I_{0} $, $i_{cir} = I_{cir}/I_{0} $,  $i_{b} = I_{b}/I_{0} $, $\phi_{a} = \Phi_{a}/\Phi_{0} $, $\epsilon = E/\frac{1}{2}L_{k}I_{0}^{2}$, $\beta_{L} = L_{t}I_{0}/\Phi_{0} = 2L_{t}I_{c}/\Phi_{0}$ and $\kappa = L_{k}/  L_{t}$. $\beta_{L}$ is the well-known screening parameter and $\kappa$ is the kinetic inductance fraction: $0 \le \kappa \le 1$. Here, instead of $L_{l}/  L_{k}$, I have preferred to express energy in terms of $\kappa$, as this is more commonly used in literature (see, e.g., Ref.\cite{day2003broadband} and references therein).

\section{Results, Analysis and Discussion}
\subsection{Variation of $i_{cs}$ and $\epsilon$ as a function of $\phi_{a}$}

\begin{figure}\centerline{\includegraphics[width=9cm,angle=0]{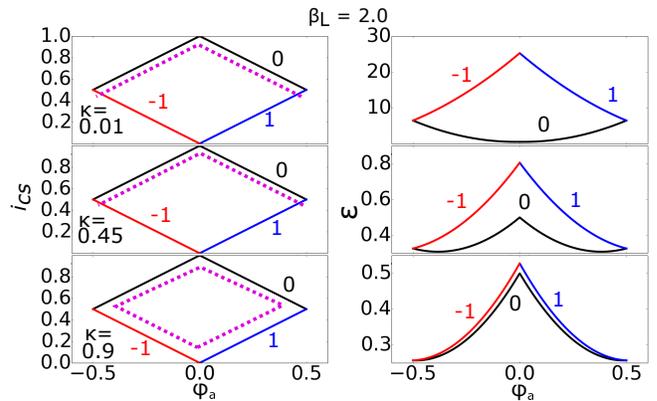}}
	\caption {Left panel: The normalized critical current, $i_{cs}$, of a NBS as a function of normalized flux, $\phi_{a}$, for three different kinetic inductance fraction, $\kappa$. All possible $i_{cs}$ branches, corresponding to different allowed fluxoid number, $n$, as per Eq.\ref{eq:nn}, are shown. The values of $n$ are represented by different colours: black (0), red (-1), and blue (1), and also indicated in the figures. The expected experimental $i_{cs}$s are indicated by dashed lines. All three curves are for the same screening parameter, $\beta_{L} = 2.0$. Right panel corresponding to the normalized energy, $\epsilon$, for the identical parameters of the left panel.}
	\label{fig:ics1}
\end{figure}

In this section, first, I analyze the variation of $i_{cs}(\phi_{a})$ and $\epsilon(\phi_{a})$, for different values of $\beta_{L}$ and $\kappa$. In Fig.\ref{fig:ics1}, I show the variation of $i_{cs}(\phi_{a})$ and $\epsilon(\phi_{a})$  for $\beta_{L} = 2.0$ and for three different $\kappa$. Since, $I_{cs}$ is periodic in $\Phi_{0}$, i.e,  $i_{cs}$ is periodic in 1, I restrict myself in the range $-0.5 \le \phi_{a} \le 0.5$. For this particular $\beta_{L}$, Eq.\ref{eq:nn} suggests that the allowed $n$ are $n = 0$ for the entire range of $\phi_{a}$: $-0.5 \le \phi_{a} \le 0.5$, and 1 and -1 for positive and negative flux axis, respectively. The corresponding $i_{cs}$ are plotted in different colours, as indicated in the figure, by solid lines. For this particular $\beta_{L}$, therefore, maximum two $I_c$ branches are possible. Out of these two, to understand, whether only one or both should be observable in an experiment, I also plot corresponding $\epsilon$ on the right-hand panel--- keeping in mind that the probability to occupy the lowest energy branch is more than the higher one. For a given $\phi_{a}$, to determine the threshold energy difference, $\Delta \epsilon_{th}$, between two branches, below which both the $I_{cs}$ branches should be experimentally observable, one requires a detailed thermodynamical analysis, which is not the aim of this article. Instead, first, I shall analyze the expected experimental $i_{cs}$ ($\phi_{a}$) qualitatively and subsequently discuss whether a single- or double-branched $i_{cs}(\phi_{a})$ would appear for an arbitrarily chosen $\Delta \epsilon_{th}$ quantitatively . 

Returning to Fig.\ref{fig:ics1}, for $\kappa = 0.01$ and $0.45$, we see that the energy is always much smaller for $n=0$ in comparison to $n=1$ and $-1$, except at the boundary: $\phi_{a} = \pm 0.5$. Thus, in this case, the probability of $n = 0$ configuration is much more than $n=1$ and $-1$ for the entire range of $-0.5 < \phi_{a} < 0.5$. Thus, in $i_{cs}$($\phi_{a}$), experimentally, only  $n = 0$ branch should be observable, with maxima at $\phi_{a} = 0$, as has been observed quite commonly in several experiments, for instances, in Refs. \cite{hasselbach2000microsquid,faucher2002niobium, hasselbach2002micro,hutchinson2003hot,hutchinson2004controlled,hutchinson2004fabrication, mandal2011diamond,hazra2013nano,russo2012nanoparticle, hazra2014nano,paul2016micron,mccaughan2016nanosquid, wu2017measurement, biswas2017josephson}. The above scenario, quite interestingly, changes  for $\kappa$ = 0.9. In this case, the energy is almost same for $n = 0$ and $+$ or $-1$.  Thus, in $i_{cs}$($\phi_{a}$), experimentally, all three $n$ --- 0,-1 and +1 --- are accesable, and $i_{cs}$ ($\phi_{a}$) should look like a diamond-shaped, as has been observed, for instances, in Refs. \cite{faucher2002niobium, hasselbach2002micro,hazra2013nano,russo2016nanosquids}. We note that the energy difference, $\Delta \epsilon_{th}$, between two branches becomes smaller and smaller as we move from center, i.e., at $\phi_{a} = 0$, towards the  edges, i.e., $\phi_{a} = \pm 0.5$. Thus, the possibility of double-valued $i_{cs}$($\phi_{a}$) near $\phi_{a} = \pm 0.5$ is more than near $\phi_{a} = 0$, leading to an incomplete-diamond-shaped $i_{cs}$($\phi_{a}$), as has been observed, for instances, in Refs. \cite{hasselbach2000microsquid,hutchinson2003hot,hutchinson2004controlled,hutchinson2004fabrication, hazra2014nano}.                 

\begin{figure}\centerline{\includegraphics[width=9cm,angle=0]{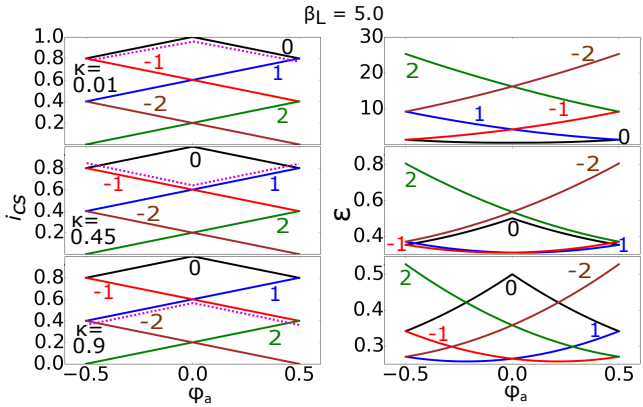}}
	\caption {Left panel: The normalized critical current, $i_{cs}$, of a NBS as a function of normalized flux, $\phi_{a}$, for different kinetic inductance fraction, $\kappa$. All possible $i_{cs}$ branches, corresponding to different allowed fluxoid number, $n$, as per Eq.\ref{eq:nn}, are shown. The values of $n$ are represented by different colours: black (0), red (-1), blue (1), green (2) and brown (-2), and also indicated in the figures. The expected experimental $i_{cs}$s are indicated schematically by dashed lines. All three curves are for the same screening parameter, $\beta_{L} = 5.0$. Right panel corresponding to the normalized energy, $\epsilon$, for the identical parameters of the left panel.}
	\label{fig:ics2}
\end{figure}

With increasing  $\beta_{L}$, more features appear. In Fig.\ref{fig:ics2}, I show the variation of $i_{cs}(\phi_{a})$ and $\epsilon(\phi_{a})$ for $\beta_{L} = 5.0$ for three different $\kappa$, identical to ones used in Fig.\ref{fig:ics1}. For this particular $\beta_{L}$, the allowed $n$ are $ 0$, $\pm 1$ and $\pm 2$ for the entire range of $\phi_{a}$. Thus, as the figures indicate, five $i_{cs}$($\phi_{a}$) branches are possible, in principle. Here, I would like to mention that experimentally, with best of my knowledge, more than two branches of $i_{cs}$($\phi_{a}$) has never been observed in NBSs \cite{granata2016nano}. This indicates that the probability to occupy the third or any of the higher branches is very small. Following the discussion of the previous paragraph, i.e., $\beta_{L} = 2.0 $ case, here also, we can qualitatively understand whether single or two branches of $i_{cs}$($\phi_{a}$) is likely to be observed in experiments. Instead, I shall discuss the other important salient features, assuming that only single-branched $i_{cs}$($\phi_{a}$), corresponding to the minimum energy, is observable. For $\kappa = 0.01$, $n=0$ corresponds to minimum energy and accordingly we get a $i_{cs}$($\phi_{a}$) with maxima at $\phi_{a} = 0$. The scenario changes quite dramatically for $\kappa = 0.45$. In this case, $n = 1$ and  $-1$  correspond to minimum energy for positive and negative flux axis, respectively. Accordingly, we get a single-branched $i_{cs}$($\phi_{a}$) with minima at $\phi_{a} = 0$. So, we see that, even for a symmetric NBS, $\phi_{a} = 0$ can correspond to minima of $i_{cs}$. This has been experimentally observed, for instances, in Refs. \cite{paul2016micron,russo2016nanosquids,hazra2013nano}. The scenario turns even more dramatic for $\kappa = 0.9$. Here, like $\kappa = 0.45$, the minimum energy is governed by $n =\pm 1$; but, $n = -1$ corresponds to minimum energy for the positive flux axis whereas $n = 1$ corresponds to minimum energy for the negative flux axis. Accordingly, we get a single-branched $i_{cs}$($\phi_{a}$) with maxima at $\phi_{a} = 0$.  It, therefore, recovers the $i_{cs}$($\phi_{a}$) pattern of $\kappa = 0.01$ case, despite the fact that differnt $n$ are stabilized in these two cases. 

\subsection{Determining whether single- or double-branched $i_{cs}$($\phi_{a}$) should be observable}

\begin{figure}\centerline{\includegraphics[width=9cm,angle=0]{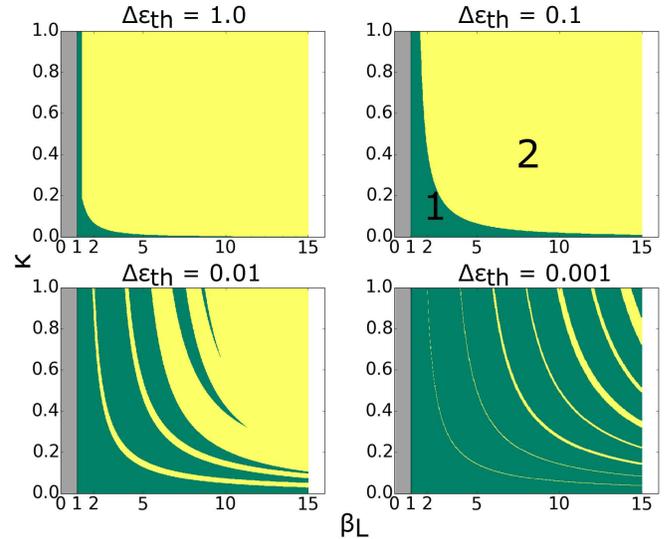}}
	\caption {The possibility of single- or double-branched $i_{cs}$($\phi_{a}$) of NBSs for different choices of threshold energy, $\Delta \epsilon_{th}$. The green colour represents single-branched (also indicated by 1) whereas the yellow color represents double-branched (also indicated by 2) $i_{cs}$($\phi_{a}$). The grey area, $\beta_{L} < 1$, is within Josephson limit and yields single-branched $i_{cs}$($\phi_{a}$); see the Discussion section in the main text for the details.}
	\label{fig:branch}
\end{figure}

From Fig.\ref{fig:ics1} and \ref{fig:ics2}, it is apparent that depending upon the values of $\beta_{L}$ and $\kappa$, $i_{cs}$($\phi_{a}$) can be single- or double-branched. In this section, I determine which combinations of $\beta_{L}$ and $\kappa$ yield single-branched and which ones yield double-branched $i_{cs}$($\phi_{a}$). To do so, I calculate the energy difference, $\Delta \epsilon$, between the first two branches, close to the edge (i.e., $\phi_{a} = \pm 0.5$ ), at an arbitrarily chosen $\phi_{a} = \pm 0.35$. I assume that $\Delta \epsilon \le \Delta \epsilon_{th} $ yields double-branched, otherwise it leads to single-branched $i_{cs}$($\phi_{a}$). In Fig.\ref{fig:branch}, I show the possibility of single- or double-branched $i_{cs}$($\phi_{a}$) for four different choices of $\Delta \epsilon_{th} $ --- 1.0, 0.1, 0.01, and 0.001 --- respectively, as a function of $\beta_{L}$ and $\kappa$. We see that for $\beta_{L} \to 1 $, irrespective of the values of $\kappa$ and for $\kappa \to 0$, irrespective of the values of $\beta_{L}$, yield single-branched $i_{cs}$($\phi_{a}$), independent to the choices of $\Delta \epsilon_{th} $. For $\Delta \epsilon_{th} $ = 1 and 0.1, at a fixed $\beta_{L}$, higher $\kappa$ values increase the probability of double-branched $i_{cs}$($\phi_{a}$). In these cases, the most of the area in the $\beta_{L}$--$\kappa$ space favours the double-branched $i_{cs}$($\phi_{a}$). With decreasing $\Delta \epsilon_{th} $, $\beta_{L}$--$\kappa$ space is devided into different domains: certain combinations of $\beta_{L}$ and $\kappa$ favours single- and the remaining combinations favours double-branched $i_{cs}$($\phi_{a}$), as expected. Furthermore, with decreasing $\Delta \epsilon_{th} $, more and more area of $\beta_{L}$--$\kappa$ space favours single-branched $i_{cs}$($\phi_{a}$). We also note that with increasing $\beta_{L}$ and $\kappa$, the area of the double-branched $i_{cs}$($\phi_{a}$) domains increases.  For materials with higher $\kappa$, like niobium and niobium nitride \cite{annunziata2010tunable}, typically also have higher critical current density compared to materials with lower $\kappa$, for instance, Al.  Thus, for identical nano-SQUID geometries, $\beta_{L}$ is also higher for high-$\kappa$ materials, making the appearance of double-branched $i_{cs}$($\phi_{a}$) more probable compared to low-$\kappa$ ones, as has been reported in several publications, for instances, in Refs. \cite{hasselbach2000microsquid,faucher2002niobium, hasselbach2002micro, hazra2014nano,russo2016nanosquids}.                          

\subsection{Calculating modulation depth and transfer function}

In this section, I shall calculate two important parameters, namely, the modulation depth and the transfer function. For simplicity, first, let me consider the case where only $n = 0$ is accessible. From Eq.\ref{eq:icss}, it is clear that maximum $i_{cs}$, $i_{cs}^{max}$, corresponds to minimum $|i_{cir}|$ whereas, minimum $i_{cs}$, $i_{cs}^{min}$, corresponds to maximum $|i_{cir}|$. For $n = 0$, Eq.\ref{eq:fqq} tells that minimum $|i_{cir}|$ is 0 whereas maximum $|i_{cir}|$ is 0.5/$\beta_{L}$ (corresponding to $\phi_{a} = $ 0 and $\pm$ 0.5, respectively). This leads $i_{cs}^{max} = 1 $ and  $i_{cs}^{min} = 1 - 1/\beta_{L}$, yielding a modulation depth 

\begin{eqnarray}
i_{cs}^{max} - i_{cs}^{min} = \frac{I_{cs}^{max} - I_{cs}^{min}}{I_{0}} = \frac{1}  {\beta_{L}},   
\label{eq:mod}
\end{eqnarray} 

in normalized unit. 

It can be shown that Eq.\ref{eq:mod} is valid in general, irrespective of whether $i_{cs}$($\phi_{a}$) is single or double-branched. This is also evident from both Fig.\ref{fig:ics1} and \ref{fig:ics2}. Here, I would like to point out that Eq.\ref{eq:mod} can be derived approximately from conventional d.c. SQUID theory \cite{clarke2006squid, granata2016nano} and has often been used in the context of NBSs.         

For the transfer function($I_{cs\Phi_{a}}$), i.e., the slope of the $I_{cs}$($\Phi_{a}$), since the variation of $I_{cs}$($\Phi_{a}$) is linear, $I_{cs\Phi_{a}}$ can straight forwardly be derived as 

\begin{eqnarray}
I_{cs\Phi_{a}} = \frac{I_{cs}^{max}-I_{cs}^{min}}{\Phi_{0}/2} = \frac{2I_{0}}{\beta_{L} \Phi_{0}}.   
\label{eq:transfun}
\end{eqnarray} 

\subsection{Limits of the model}

I have shown that using the model presented here, which does not take the Josephson effect in NB--electrode joints into account, $I_{cs}$($\Phi_{a}$) of NBSs is derivable. The result is triangular-shaped $I_{cs}$($\Phi_{a}$) with one to two branches, as has been observed in several experiments \cite{hasselbach2000microsquid,faucher2002niobium, hasselbach2002micro,hutchinson2003hot,hutchinson2004controlled,hutchinson2004fabrication, mandal2011diamond,hazra2013nano,russo2012nanoparticle, hazra2014nano,paul2016micron,mccaughan2016nanosquid, wu2017measurement, biswas2017josephson}. Now, let me discuss the limits in which the model works. The central assumption of the model is that the phase drop across the NBs is not significantly higher than the overall phase drop across the electrodes of the NBS. This assumption is valid for NBs longer than $\xi_{T}$. A large number of NBSs reported in the literature fulfils this criterion (see for instance Ref.\cite{granata2016nano} and references therein). As the length of the NBs approaches $\xi_{T}$, a well-defined $\theta$ can be attributed to the NBs and they approximately behave like Josephson junctions \cite{likharev1979superconducting,tinkham1996introduction}--- consequently, $I_{cs}$($\Phi_{a}$) of a NBS deviates from being triangular and becomes more sinusoidal \cite{hasselbach2002micro,vijay2010approaching}. Another restriction comes from Eq.\ref{eq:nn} which imposes that $\beta_{L}$ must be  $\ge 1.0$. Like a short NB, for $\beta_{L} \le 1.0$ also, a NB behaves more like a Josephson junction.

\section{Conclusion}

In conclusion, I have developed a model for NBSs beyond the Josephson limit, i.e., for long NBs and/or large screening parameter. In this limit, the $I_{cs}$($\Phi_{a}$) of a NBS can be understood by considering the fluxoid quantization in the NBS loop and the energy of the NBS. The model explains various experimental features --- like, triangular-shaped, double-branched, and a diamond-shaped $I_{c}$ ($\Phi_{a}$) --- reported in the literature. From the model, I derive the expression for the modulation depth and the transfer function. Using the model, I have shown that both the screening parameter and the kinetic inductance fraction play vital role in deciding the number of $I_{cs}$($\Phi_{a}$) branches.     

\section{Acknowledgements}

I acknowledge the financial support from the CSIR India.

\bibliography{Bibliography}

\begin{thebibliography}{50}%
\makeatletter
\providecommand \@ifxundefined [1]{%
 \@ifx{#1\undefined}
}%
\providecommand \@ifnum [1]{%
 \ifnum #1\expandafter \@firstoftwo
 \else \expandafter \@secondoftwo
 \fi
}%
\providecommand \@ifx [1]{%
 \ifx #1\expandafter \@firstoftwo
 \else \expandafter \@secondoftwo
 \fi
}%
\providecommand \natexlab [1]{#1}%
\providecommand \enquote  [1]{``#1''}%
\providecommand \bibnamefont  [1]{#1}%
\providecommand \bibfnamefont [1]{#1}%
\providecommand \citenamefont [1]{#1}%
\providecommand \href@noop [0]{\@secondoftwo}%
\providecommand \href [0]{\begingroup \@sanitize@url \@href}%
\providecommand \@href[1]{\@@startlink{#1}\@@href}%
\providecommand \@@href[1]{\endgroup#1\@@endlink}%
\providecommand \@sanitize@url [0]{\catcode `\\12\catcode `\$12\catcode
  `\&12\catcode `\#12\catcode `\^12\catcode `\_12\catcode `\%12\relax}%
\providecommand \@@startlink[1]{}%
\providecommand \@@endlink[0]{}%
\providecommand \url  [0]{\begingroup\@sanitize@url \@url }%
\providecommand \@url [1]{\endgroup\@href {#1}{\urlprefix }}%
\providecommand \urlprefix  [0]{URL }%
\providecommand \Eprint [0]{\href }%
\providecommand \doibase [0]{http://dx.doi.org/}%
\providecommand \selectlanguage [0]{\@gobble}%
\providecommand \bibinfo  [0]{\@secondoftwo}%
\providecommand \bibfield  [0]{\@secondoftwo}%
\providecommand \translation [1]{[#1]}%
\providecommand \BibitemOpen [0]{}%
\providecommand \bibitemStop [0]{}%
\providecommand \bibitemNoStop [0]{.\EOS\space}%
\providecommand \EOS [0]{\spacefactor3000\relax}%
\providecommand \BibitemShut  [1]{\csname bibitem#1\endcsname}%
\let\auto@bib@innerbib\@empty
%</preamble>
\bibitem [{\citenamefont {Wernsdorfer}(2007)}]{wernsdorfer2007classical}%
  \BibitemOpen
  \bibfield  {author} {\bibinfo {author} {\bibfnamefont {W.}~\bibnamefont
  {Wernsdorfer}},\ }\href@noop {} {\bibfield  {journal} {\bibinfo  {journal}
  {Advances in Chemical Physics, Volume 118}\ ,\ \bibinfo {pages} {99}}
  (\bibinfo {year} {2007})}\BibitemShut {NoStop}%
\bibitem [{\citenamefont {Wernsdorfer}(2009)}]{wernsdorfer2009micro}%
  \BibitemOpen
  \bibfield  {author} {\bibinfo {author} {\bibfnamefont {W.}~\bibnamefont
  {Wernsdorfer}},\ }\href@noop {} {\bibfield  {journal} {\bibinfo  {journal}
  {Superconductor Science and Technology}\ }\textbf {\bibinfo {volume} {22}},\
  \bibinfo {pages} {064013} (\bibinfo {year} {2009})}\BibitemShut {NoStop}%
\bibitem [{\citenamefont {Foley}\ and\ \citenamefont
  {Hilgenkamp}(2009)}]{foley2009nanosquids}%
  \BibitemOpen
  \bibfield  {author} {\bibinfo {author} {\bibfnamefont {C.}~\bibnamefont
  {Foley}}\ and\ \bibinfo {author} {\bibfnamefont {H.}~\bibnamefont
  {Hilgenkamp}},\ }\href@noop {} {\bibfield  {journal} {\bibinfo  {journal}
  {Superconductor science and technology}\ }\textbf {\bibinfo {volume} {22}},\
  \bibinfo {pages} {064001} (\bibinfo {year} {2009})}\BibitemShut {NoStop}%
\bibitem [{\citenamefont {Vasyukov}\ \emph {et~al.}(2013)\citenamefont
  {Vasyukov}, \citenamefont {Anahory}, \citenamefont {Embon}, \citenamefont
  {Halbertal}, \citenamefont {Cuppens}, \citenamefont {Neeman}, \citenamefont
  {Finkler}, \citenamefont {Segev}, \citenamefont {Myasoedov}, \citenamefont
  {Rappaport} \emph {et~al.}}]{vasyukov2013scanning}%
  \BibitemOpen
  \bibfield  {author} {\bibinfo {author} {\bibfnamefont {D.}~\bibnamefont
  {Vasyukov}}, \bibinfo {author} {\bibfnamefont {Y.}~\bibnamefont {Anahory}},
  \bibinfo {author} {\bibfnamefont {L.}~\bibnamefont {Embon}}, \bibinfo
  {author} {\bibfnamefont {D.}~\bibnamefont {Halbertal}}, \bibinfo {author}
  {\bibfnamefont {J.}~\bibnamefont {Cuppens}}, \bibinfo {author} {\bibfnamefont
  {L.}~\bibnamefont {Neeman}}, \bibinfo {author} {\bibfnamefont
  {A.}~\bibnamefont {Finkler}}, \bibinfo {author} {\bibfnamefont
  {Y.}~\bibnamefont {Segev}}, \bibinfo {author} {\bibfnamefont
  {Y.}~\bibnamefont {Myasoedov}}, \bibinfo {author} {\bibfnamefont {M.~L.}\
  \bibnamefont {Rappaport}},  \emph {et~al.},\ }\href@noop {} {\bibfield
  {journal} {\bibinfo  {journal} {Nature nanotechnology}\ }\textbf {\bibinfo
  {volume} {8}},\ \bibinfo {pages} {639} (\bibinfo {year} {2013})}\BibitemShut
  {NoStop}%
\bibitem [{\citenamefont {Levenson-Falk}\ \emph {et~al.}(2013)\citenamefont
  {Levenson-Falk}, \citenamefont {Vijay}, \citenamefont {Antler},\ and\
  \citenamefont {Siddiqi}}]{levenson2013dispersive}%
  \BibitemOpen
  \bibfield  {author} {\bibinfo {author} {\bibfnamefont {E.}~\bibnamefont
  {Levenson-Falk}}, \bibinfo {author} {\bibfnamefont {R.}~\bibnamefont
  {Vijay}}, \bibinfo {author} {\bibfnamefont {N.}~\bibnamefont {Antler}}, \
  and\ \bibinfo {author} {\bibfnamefont {I.}~\bibnamefont {Siddiqi}},\
  }\href@noop {} {\bibfield  {journal} {\bibinfo  {journal} {Superconductor
  Science and Technology}\ }\textbf {\bibinfo {volume} {26}},\ \bibinfo {pages}
  {055015} (\bibinfo {year} {2013})}\BibitemShut {NoStop}%
\bibitem [{\citenamefont {Granata}\ and\ \citenamefont
  {Vettoliere}(2016)}]{granata2016nano}%
  \BibitemOpen
  \bibfield  {author} {\bibinfo {author} {\bibfnamefont {C.}~\bibnamefont
  {Granata}}\ and\ \bibinfo {author} {\bibfnamefont {A.}~\bibnamefont
  {Vettoliere}},\ }\href@noop {} {\bibfield  {journal} {\bibinfo  {journal}
  {Physics Reports}\ }\textbf {\bibinfo {volume} {614}},\ \bibinfo {pages} {1}
  (\bibinfo {year} {2016})}\BibitemShut {NoStop}%
\bibitem [{\citenamefont {Gallop}\ and\ \citenamefont
  {Hao}(2016)}]{gallop2016nanoscale}%
  \BibitemOpen
  \bibfield  {author} {\bibinfo {author} {\bibfnamefont {J.}~\bibnamefont
  {Gallop}}\ and\ \bibinfo {author} {\bibfnamefont {L.}~\bibnamefont {Hao}},\
  }\href@noop {} {\bibfield  {journal} {\bibinfo  {journal} {ACS nano}\
  }\textbf {\bibinfo {volume} {10}},\ \bibinfo {pages} {8128} (\bibinfo {year}
  {2016})}\BibitemShut {NoStop}%
\bibitem [{\citenamefont {Yue}\ \emph {et~al.}(2017)\citenamefont {Yue},
  \citenamefont {Chen}, \citenamefont {Barreda}, \citenamefont {Bevara},
  \citenamefont {Hu}, \citenamefont {Wu}, \citenamefont {Wang}, \citenamefont
  {Andrei}, \citenamefont {Bertaina},\ and\ \citenamefont
  {Chiorescu}}]{yue2017sensitive}%
  \BibitemOpen
  \bibfield  {author} {\bibinfo {author} {\bibfnamefont {G.}~\bibnamefont
  {Yue}}, \bibinfo {author} {\bibfnamefont {L.}~\bibnamefont {Chen}}, \bibinfo
  {author} {\bibfnamefont {J.}~\bibnamefont {Barreda}}, \bibinfo {author}
  {\bibfnamefont {V.}~\bibnamefont {Bevara}}, \bibinfo {author} {\bibfnamefont
  {L.}~\bibnamefont {Hu}}, \bibinfo {author} {\bibfnamefont {L.}~\bibnamefont
  {Wu}}, \bibinfo {author} {\bibfnamefont {Z.}~\bibnamefont {Wang}}, \bibinfo
  {author} {\bibfnamefont {P.}~\bibnamefont {Andrei}}, \bibinfo {author}
  {\bibfnamefont {S.}~\bibnamefont {Bertaina}}, \ and\ \bibinfo {author}
  {\bibfnamefont {I.}~\bibnamefont {Chiorescu}},\ }\href@noop {} {\bibfield
  {journal} {\bibinfo  {journal} {Applied Physics Letters}\ }\textbf {\bibinfo
  {volume} {111}},\ \bibinfo {pages} {202601} (\bibinfo {year}
  {2017})}\BibitemShut {NoStop}%
\bibitem [{\citenamefont {Mailly}\ \emph {et~al.}(1993)\citenamefont {Mailly},
  \citenamefont {Chapelier},\ and\ \citenamefont
  {Benoit}}]{mailly1993experimental}%
  \BibitemOpen
  \bibfield  {author} {\bibinfo {author} {\bibfnamefont {D.}~\bibnamefont
  {Mailly}}, \bibinfo {author} {\bibfnamefont {C.}~\bibnamefont {Chapelier}}, \
  and\ \bibinfo {author} {\bibfnamefont {A.}~\bibnamefont {Benoit}},\
  }\href@noop {} {\bibfield  {journal} {\bibinfo  {journal} {Physical review
  letters}\ }\textbf {\bibinfo {volume} {70}},\ \bibinfo {pages} {2020}
  (\bibinfo {year} {1993})}\BibitemShut {NoStop}%
\bibitem [{\citenamefont {Rabaud}\ \emph {et~al.}(2001)\citenamefont {Rabaud},
  \citenamefont {Saminadayar}, \citenamefont {Mailly}, \citenamefont
  {Hasselbach}, \citenamefont {Benoit},\ and\ \citenamefont
  {Etienne}}]{rabaud2001persistent}%
  \BibitemOpen
  \bibfield  {author} {\bibinfo {author} {\bibfnamefont {W.}~\bibnamefont
  {Rabaud}}, \bibinfo {author} {\bibfnamefont {L.}~\bibnamefont {Saminadayar}},
  \bibinfo {author} {\bibfnamefont {D.}~\bibnamefont {Mailly}}, \bibinfo
  {author} {\bibfnamefont {K.}~\bibnamefont {Hasselbach}}, \bibinfo {author}
  {\bibfnamefont {A.}~\bibnamefont {Benoit}}, \ and\ \bibinfo {author}
  {\bibfnamefont {B.}~\bibnamefont {Etienne}},\ }\href@noop {} {\bibfield
  {journal} {\bibinfo  {journal} {Physical Review Letters}\ }\textbf {\bibinfo
  {volume} {86}},\ \bibinfo {pages} {3124} (\bibinfo {year}
  {2001})}\BibitemShut {NoStop}%
\bibitem [{\citenamefont {Hao}\ \emph {et~al.}(2003)\citenamefont {Hao},
  \citenamefont {Gallop}, \citenamefont {Gardiner}, \citenamefont
  {Josephs-Franks}, \citenamefont {Macfarlane}, \citenamefont {Lam},\ and\
  \citenamefont {Foley}}]{hao2003inductive}%
  \BibitemOpen
  \bibfield  {author} {\bibinfo {author} {\bibfnamefont {L.}~\bibnamefont
  {Hao}}, \bibinfo {author} {\bibfnamefont {J.}~\bibnamefont {Gallop}},
  \bibinfo {author} {\bibfnamefont {C.}~\bibnamefont {Gardiner}}, \bibinfo
  {author} {\bibfnamefont {P.}~\bibnamefont {Josephs-Franks}}, \bibinfo
  {author} {\bibfnamefont {J.}~\bibnamefont {Macfarlane}}, \bibinfo {author}
  {\bibfnamefont {S.}~\bibnamefont {Lam}}, \ and\ \bibinfo {author}
  {\bibfnamefont {C.}~\bibnamefont {Foley}},\ }\href@noop {} {\bibfield
  {journal} {\bibinfo  {journal} {Superconductor Science and Technology}\
  }\textbf {\bibinfo {volume} {16}},\ \bibinfo {pages} {1479} (\bibinfo {year}
  {2003})}\BibitemShut {NoStop}%
\bibitem [{\citenamefont {Etaki}\ \emph {et~al.}(2008)\citenamefont {Etaki},
  \citenamefont {Poot}, \citenamefont {Mahboob}, \citenamefont {Onomitsu},
  \citenamefont {Yamaguchi},\ and\ \citenamefont {Van~der
  Zant}}]{etaki2008motion}%
  \BibitemOpen
  \bibfield  {author} {\bibinfo {author} {\bibfnamefont {S.}~\bibnamefont
  {Etaki}}, \bibinfo {author} {\bibfnamefont {M.}~\bibnamefont {Poot}},
  \bibinfo {author} {\bibfnamefont {I.}~\bibnamefont {Mahboob}}, \bibinfo
  {author} {\bibfnamefont {K.}~\bibnamefont {Onomitsu}}, \bibinfo {author}
  {\bibfnamefont {H.}~\bibnamefont {Yamaguchi}}, \ and\ \bibinfo {author}
  {\bibfnamefont {H.}~\bibnamefont {Van~der Zant}},\ }\href@noop {} {\bibfield
  {journal} {\bibinfo  {journal} {Nature Physics}\ }\textbf {\bibinfo {volume}
  {4}},\ \bibinfo {pages} {785} (\bibinfo {year} {2008})}\BibitemShut {NoStop}%
\bibitem [{\citenamefont {Vijay}\ \emph {et~al.}(2009)\citenamefont {Vijay},
  \citenamefont {Sau}, \citenamefont {Cohen},\ and\ \citenamefont
  {Siddiqi}}]{vijay2009optimizing}%
  \BibitemOpen
  \bibfield  {author} {\bibinfo {author} {\bibfnamefont {R.}~\bibnamefont
  {Vijay}}, \bibinfo {author} {\bibfnamefont {J.}~\bibnamefont {Sau}}, \bibinfo
  {author} {\bibfnamefont {M.~L.}\ \bibnamefont {Cohen}}, \ and\ \bibinfo
  {author} {\bibfnamefont {I.}~\bibnamefont {Siddiqi}},\ }\href@noop {}
  {\bibfield  {journal} {\bibinfo  {journal} {Physical review letters}\
  }\textbf {\bibinfo {volume} {103}},\ \bibinfo {pages} {087003} (\bibinfo
  {year} {2009})}\BibitemShut {NoStop}%
\bibitem [{\citenamefont {Mart{\'\i}nez-P{\'e}rez}\ and\ \citenamefont
  {Koelle}(2016)}]{martinez2016nanosquids}%
  \BibitemOpen
  \bibfield  {author} {\bibinfo {author} {\bibfnamefont {M.~J.}\ \bibnamefont
  {Mart{\'\i}nez-P{\'e}rez}}\ and\ \bibinfo {author} {\bibfnamefont
  {D.}~\bibnamefont {Koelle}},\ }\href@noop {} {\bibfield  {journal} {\bibinfo
  {journal} {Physical Sciences Reviews}\ }\textbf {\bibinfo {volume} {2}}
  (\bibinfo {year} {2016})}\BibitemShut {NoStop}%
\bibitem [{\citenamefont {Hasselbach}\ \emph {et~al.}(2002)\citenamefont
  {Hasselbach}, \citenamefont {Mailly},\ and\ \citenamefont
  {Kirtley}}]{hasselbach2002micro}%
  \BibitemOpen
  \bibfield  {author} {\bibinfo {author} {\bibfnamefont {K.}~\bibnamefont
  {Hasselbach}}, \bibinfo {author} {\bibfnamefont {D.}~\bibnamefont {Mailly}},
  \ and\ \bibinfo {author} {\bibfnamefont {J.}~\bibnamefont {Kirtley}},\
  }\href@noop {} {\bibfield  {journal} {\bibinfo  {journal} {Journal of applied
  physics}\ }\textbf {\bibinfo {volume} {91}},\ \bibinfo {pages} {4432}
  (\bibinfo {year} {2002})}\BibitemShut {NoStop}%
\bibitem [{\citenamefont {Lam}\ and\ \citenamefont
  {Tilbrook}(2003)}]{lam2003development}%
  \BibitemOpen
  \bibfield  {author} {\bibinfo {author} {\bibfnamefont {S.}~\bibnamefont
  {Lam}}\ and\ \bibinfo {author} {\bibfnamefont {D.}~\bibnamefont {Tilbrook}},\
  }\href@noop {} {\bibfield  {journal} {\bibinfo  {journal} {Applied physics
  letters}\ }\textbf {\bibinfo {volume} {82}},\ \bibinfo {pages} {1078}
  (\bibinfo {year} {2003})}\BibitemShut {NoStop}%
\bibitem [{\citenamefont {Troeman}\ \emph {et~al.}(2007)\citenamefont
  {Troeman}, \citenamefont {Derking}, \citenamefont {Borger}, \citenamefont
  {Pleikies}, \citenamefont {Veldhuis},\ and\ \citenamefont
  {Hilgenkamp}}]{troeman2007nanosquids}%
  \BibitemOpen
  \bibfield  {author} {\bibinfo {author} {\bibfnamefont {A.~G.}\ \bibnamefont
  {Troeman}}, \bibinfo {author} {\bibfnamefont {H.}~\bibnamefont {Derking}},
  \bibinfo {author} {\bibfnamefont {B.}~\bibnamefont {Borger}}, \bibinfo
  {author} {\bibfnamefont {J.}~\bibnamefont {Pleikies}}, \bibinfo {author}
  {\bibfnamefont {D.}~\bibnamefont {Veldhuis}}, \ and\ \bibinfo {author}
  {\bibfnamefont {H.}~\bibnamefont {Hilgenkamp}},\ }\href@noop {} {\bibfield
  {journal} {\bibinfo  {journal} {Nano Letters}\ }\textbf {\bibinfo {volume}
  {7}},\ \bibinfo {pages} {2152} (\bibinfo {year} {2007})}\BibitemShut
  {NoStop}%
\bibitem [{\citenamefont {Hao}\ \emph {et~al.}(2008)\citenamefont {Hao},
  \citenamefont {Macfarlane}, \citenamefont {Gallop}, \citenamefont {Cox},
  \citenamefont {Beyer}, \citenamefont {Drung},\ and\ \citenamefont
  {Schurig}}]{hao2008measurement}%
  \BibitemOpen
  \bibfield  {author} {\bibinfo {author} {\bibfnamefont {L.}~\bibnamefont
  {Hao}}, \bibinfo {author} {\bibfnamefont {J.}~\bibnamefont {Macfarlane}},
  \bibinfo {author} {\bibfnamefont {J.}~\bibnamefont {Gallop}}, \bibinfo
  {author} {\bibfnamefont {D.}~\bibnamefont {Cox}}, \bibinfo {author}
  {\bibfnamefont {J.}~\bibnamefont {Beyer}}, \bibinfo {author} {\bibfnamefont
  {D.}~\bibnamefont {Drung}}, \ and\ \bibinfo {author} {\bibfnamefont
  {T.}~\bibnamefont {Schurig}},\ }\href@noop {} {\bibfield  {journal} {\bibinfo
   {journal} {Applied Physics Letters}\ }\textbf {\bibinfo {volume} {92}},\
  \bibinfo {pages} {192507} (\bibinfo {year} {2008})}\BibitemShut {NoStop}%
\bibitem [{\citenamefont {Vijay}\ \emph {et~al.}(2010)\citenamefont {Vijay},
  \citenamefont {Levenson-Falk}, \citenamefont {Slichter},\ and\ \citenamefont
  {Siddiqi}}]{vijay2010approaching}%
  \BibitemOpen
  \bibfield  {author} {\bibinfo {author} {\bibfnamefont {R.}~\bibnamefont
  {Vijay}}, \bibinfo {author} {\bibfnamefont {E.}~\bibnamefont
  {Levenson-Falk}}, \bibinfo {author} {\bibfnamefont {D.}~\bibnamefont
  {Slichter}}, \ and\ \bibinfo {author} {\bibfnamefont {I.}~\bibnamefont
  {Siddiqi}},\ }\href@noop {} {\bibfield  {journal} {\bibinfo  {journal}
  {Applied Physics Letters}\ }\textbf {\bibinfo {volume} {96}},\ \bibinfo
  {pages} {223112} (\bibinfo {year} {2010})}\BibitemShut {NoStop}%
\bibitem [{\citenamefont {Mandal}\ \emph {et~al.}(2011)\citenamefont {Mandal},
  \citenamefont {Bautze}, \citenamefont {Williams}, \citenamefont {Naud},
  \citenamefont {Bustarret}, \citenamefont {Omnes}, \citenamefont {Rodiere},
  \citenamefont {Meunier}, \citenamefont {Bäuerle},\ and\ \citenamefont
  {Saminadayar}}]{mandal2011diamond}%
  \BibitemOpen
  \bibfield  {author} {\bibinfo {author} {\bibfnamefont {S.}~\bibnamefont
  {Mandal}}, \bibinfo {author} {\bibfnamefont {T.}~\bibnamefont {Bautze}},
  \bibinfo {author} {\bibfnamefont {O.~A.}\ \bibnamefont {Williams}}, \bibinfo
  {author} {\bibfnamefont {C.}~\bibnamefont {Naud}}, \bibinfo {author}
  {\bibfnamefont {E.}~\bibnamefont {Bustarret}}, \bibinfo {author}
  {\bibfnamefont {F.}~\bibnamefont {Omnes}}, \bibinfo {author} {\bibfnamefont
  {P.}~\bibnamefont {Rodiere}}, \bibinfo {author} {\bibfnamefont
  {T.}~\bibnamefont {Meunier}}, \bibinfo {author} {\bibfnamefont
  {C.}~\bibnamefont {Bäuerle}}, \ and\ \bibinfo {author} {\bibfnamefont
  {L.}~\bibnamefont {Saminadayar}},\ }\href@noop {} {\bibfield  {journal}
  {\bibinfo  {journal} {ACS nano}\ }\textbf {\bibinfo {volume} {5}},\ \bibinfo
  {pages} {7144} (\bibinfo {year} {2011})}\BibitemShut {NoStop}%
\bibitem [{\citenamefont {Hazra}\ \emph {et~al.}(2013)\citenamefont {Hazra},
  \citenamefont {Kirtley},\ and\ \citenamefont {Hasselbach}}]{hazra2013nano}%
  \BibitemOpen
  \bibfield  {author} {\bibinfo {author} {\bibfnamefont {D.}~\bibnamefont
  {Hazra}}, \bibinfo {author} {\bibfnamefont {J.~R.}\ \bibnamefont {Kirtley}},
  \ and\ \bibinfo {author} {\bibfnamefont {K.}~\bibnamefont {Hasselbach}},\
  }\href@noop {} {\bibfield  {journal} {\bibinfo  {journal} {Applied Physics
  Letters}\ }\textbf {\bibinfo {volume} {103}},\ \bibinfo {pages} {093109}
  (\bibinfo {year} {2013})}\BibitemShut {NoStop}%
\bibitem [{\citenamefont {Chen}\ \emph {et~al.}(2016)\citenamefont {Chen},
  \citenamefont {Wang}, \citenamefont {Liu}, \citenamefont {Wu},\ and\
  \citenamefont {Wang}}]{chen2016high}%
  \BibitemOpen
  \bibfield  {author} {\bibinfo {author} {\bibfnamefont {L.}~\bibnamefont
  {Chen}}, \bibinfo {author} {\bibfnamefont {H.}~\bibnamefont {Wang}}, \bibinfo
  {author} {\bibfnamefont {X.}~\bibnamefont {Liu}}, \bibinfo {author}
  {\bibfnamefont {L.}~\bibnamefont {Wu}}, \ and\ \bibinfo {author}
  {\bibfnamefont {Z.}~\bibnamefont {Wang}},\ }\href@noop {} {\bibfield
  {journal} {\bibinfo  {journal} {Nano letters}\ }\textbf {\bibinfo {volume}
  {16}},\ \bibinfo {pages} {7726} (\bibinfo {year} {2016})}\BibitemShut
  {NoStop}%
\bibitem [{\citenamefont {Angers}\ \emph {et~al.}(2008)\citenamefont {Angers},
  \citenamefont {Chiodi}, \citenamefont {Montambaux}, \citenamefont {Ferrier},
  \citenamefont {Gu{\'e}ron}, \citenamefont {Bouchiat},\ and\ \citenamefont
  {Cuevas}}]{angers2008proximity}%
  \BibitemOpen
  \bibfield  {author} {\bibinfo {author} {\bibfnamefont {L.}~\bibnamefont
  {Angers}}, \bibinfo {author} {\bibfnamefont {F.}~\bibnamefont {Chiodi}},
  \bibinfo {author} {\bibfnamefont {G.}~\bibnamefont {Montambaux}}, \bibinfo
  {author} {\bibfnamefont {M.}~\bibnamefont {Ferrier}}, \bibinfo {author}
  {\bibfnamefont {S.}~\bibnamefont {Gu{\'e}ron}}, \bibinfo {author}
  {\bibfnamefont {H.}~\bibnamefont {Bouchiat}}, \ and\ \bibinfo {author}
  {\bibfnamefont {J.}~\bibnamefont {Cuevas}},\ }\href@noop {} {\bibfield
  {journal} {\bibinfo  {journal} {Physical Review B}\ }\textbf {\bibinfo
  {volume} {77}},\ \bibinfo {pages} {165408} (\bibinfo {year}
  {2008})}\BibitemShut {NoStop}%
\bibitem [{\citenamefont {Ronzani}\ \emph {et~al.}(2013)\citenamefont
  {Ronzani}, \citenamefont {Baillergeau}, \citenamefont {Altimiras},\ and\
  \citenamefont {Giazotto}}]{ronzani2013micro}%
  \BibitemOpen
  \bibfield  {author} {\bibinfo {author} {\bibfnamefont {A.}~\bibnamefont
  {Ronzani}}, \bibinfo {author} {\bibfnamefont {M.}~\bibnamefont
  {Baillergeau}}, \bibinfo {author} {\bibfnamefont {C.}~\bibnamefont
  {Altimiras}}, \ and\ \bibinfo {author} {\bibfnamefont {F.}~\bibnamefont
  {Giazotto}},\ }\href@noop {} {\bibfield  {journal} {\bibinfo  {journal}
  {Applied Physics Letters}\ }\textbf {\bibinfo {volume} {103}},\ \bibinfo
  {pages} {052603} (\bibinfo {year} {2013})}\BibitemShut {NoStop}%
\bibitem [{\citenamefont {Samaddar}\ \emph {et~al.}(2013)\citenamefont
  {Samaddar}, \citenamefont {Van~Zanten}, \citenamefont {Fay}, \citenamefont
  {Sac{\'e}p{\'e}}, \citenamefont {Courtois},\ and\ \citenamefont
  {Winkelmann}}]{samaddar2013niobium}%
  \BibitemOpen
  \bibfield  {author} {\bibinfo {author} {\bibfnamefont {S.}~\bibnamefont
  {Samaddar}}, \bibinfo {author} {\bibfnamefont {D.}~\bibnamefont
  {Van~Zanten}}, \bibinfo {author} {\bibfnamefont {A.}~\bibnamefont {Fay}},
  \bibinfo {author} {\bibfnamefont {B.}~\bibnamefont {Sac{\'e}p{\'e}}},
  \bibinfo {author} {\bibfnamefont {H.}~\bibnamefont {Courtois}}, \ and\
  \bibinfo {author} {\bibfnamefont {C.}~\bibnamefont {Winkelmann}},\
  }\href@noop {} {\bibfield  {journal} {\bibinfo  {journal} {Nanotechnology}\
  }\textbf {\bibinfo {volume} {24}},\ \bibinfo {pages} {375304} (\bibinfo
  {year} {2013})}\BibitemShut {NoStop}%
\bibitem [{\citenamefont {W{\"o}lbing}\ \emph {et~al.}(2013)\citenamefont
  {W{\"o}lbing}, \citenamefont {Nagel}, \citenamefont {Schwarz}, \citenamefont
  {Kieler}, \citenamefont {Weimann}, \citenamefont {Kohlmann}, \citenamefont
  {Zorin}, \citenamefont {Kemmler}, \citenamefont {Kleiner},\ and\
  \citenamefont {Koelle}}]{wolbing2013nb}%
  \BibitemOpen
  \bibfield  {author} {\bibinfo {author} {\bibfnamefont {R.}~\bibnamefont
  {W{\"o}lbing}}, \bibinfo {author} {\bibfnamefont {J.}~\bibnamefont {Nagel}},
  \bibinfo {author} {\bibfnamefont {T.}~\bibnamefont {Schwarz}}, \bibinfo
  {author} {\bibfnamefont {O.}~\bibnamefont {Kieler}}, \bibinfo {author}
  {\bibfnamefont {T.}~\bibnamefont {Weimann}}, \bibinfo {author} {\bibfnamefont
  {J.}~\bibnamefont {Kohlmann}}, \bibinfo {author} {\bibfnamefont
  {A.}~\bibnamefont {Zorin}}, \bibinfo {author} {\bibfnamefont
  {M.}~\bibnamefont {Kemmler}}, \bibinfo {author} {\bibfnamefont
  {R.}~\bibnamefont {Kleiner}}, \ and\ \bibinfo {author} {\bibfnamefont
  {D.}~\bibnamefont {Koelle}},\ }\href@noop {} {\bibfield  {journal} {\bibinfo
  {journal} {Applied Physics Letters}\ }\textbf {\bibinfo {volume} {102}},\
  \bibinfo {pages} {192601} (\bibinfo {year} {2013})}\BibitemShut {NoStop}%
\bibitem [{\citenamefont {Granata}\ \emph {et~al.}(2013)\citenamefont
  {Granata}, \citenamefont {Vettoliere}, \citenamefont {Russo}, \citenamefont
  {Fretto}, \citenamefont {De~Leo},\ and\ \citenamefont
  {Lacquaniti}}]{granata2013three}%
  \BibitemOpen
  \bibfield  {author} {\bibinfo {author} {\bibfnamefont {C.}~\bibnamefont
  {Granata}}, \bibinfo {author} {\bibfnamefont {A.}~\bibnamefont {Vettoliere}},
  \bibinfo {author} {\bibfnamefont {R.}~\bibnamefont {Russo}}, \bibinfo
  {author} {\bibfnamefont {M.}~\bibnamefont {Fretto}}, \bibinfo {author}
  {\bibfnamefont {N.}~\bibnamefont {De~Leo}}, \ and\ \bibinfo {author}
  {\bibfnamefont {V.}~\bibnamefont {Lacquaniti}},\ }\href@noop {} {\bibfield
  {journal} {\bibinfo  {journal} {Applied Physics Letters}\ }\textbf {\bibinfo
  {volume} {103}},\ \bibinfo {pages} {102602} (\bibinfo {year}
  {2013})}\BibitemShut {NoStop}%
\bibitem [{\citenamefont {Schmelz}\ \emph {et~al.}(2016)\citenamefont
  {Schmelz}, \citenamefont {Zakosarenko}, \citenamefont {Sch{\"o}nau},
  \citenamefont {Anders}, \citenamefont {Linzen}, \citenamefont {Stolz},\ and\
  \citenamefont {Meyer}}]{schmelz2016nearly}%
  \BibitemOpen
  \bibfield  {author} {\bibinfo {author} {\bibfnamefont {M.}~\bibnamefont
  {Schmelz}}, \bibinfo {author} {\bibfnamefont {V.}~\bibnamefont
  {Zakosarenko}}, \bibinfo {author} {\bibfnamefont {T.}~\bibnamefont
  {Sch{\"o}nau}}, \bibinfo {author} {\bibfnamefont {S.}~\bibnamefont {Anders}},
  \bibinfo {author} {\bibfnamefont {S.}~\bibnamefont {Linzen}}, \bibinfo
  {author} {\bibfnamefont {R.}~\bibnamefont {Stolz}}, \ and\ \bibinfo {author}
  {\bibfnamefont {H.}~\bibnamefont {Meyer}},\ }\href@noop {} {\bibfield
  {journal} {\bibinfo  {journal} {Superconductor Science and Technology}\
  }\textbf {\bibinfo {volume} {30}},\ \bibinfo {pages} {014001} (\bibinfo
  {year} {2016})}\BibitemShut {NoStop}%
\bibitem [{\citenamefont {Cleuziou}\ \emph {et~al.}(2006)\citenamefont
  {Cleuziou}, \citenamefont {Wernsdorfer}, \citenamefont {Bouchiat},
  \citenamefont {Ondar{\c{c}}uhu},\ and\ \citenamefont
  {Monthioux}}]{cleuziou2006carbon}%
  \BibitemOpen
  \bibfield  {author} {\bibinfo {author} {\bibfnamefont {J.-P.}\ \bibnamefont
  {Cleuziou}}, \bibinfo {author} {\bibfnamefont {W.}~\bibnamefont
  {Wernsdorfer}}, \bibinfo {author} {\bibfnamefont {V.}~\bibnamefont
  {Bouchiat}}, \bibinfo {author} {\bibfnamefont {T.}~\bibnamefont
  {Ondar{\c{c}}uhu}}, \ and\ \bibinfo {author} {\bibfnamefont {M.}~\bibnamefont
  {Monthioux}},\ }\href@noop {} {\bibfield  {journal} {\bibinfo  {journal}
  {Nature nanotechnology}\ }\textbf {\bibinfo {volume} {1}},\ \bibinfo {pages}
  {53} (\bibinfo {year} {2006})}\BibitemShut {NoStop}%
\bibitem [{\citenamefont {Clarke}\ and\ \citenamefont
  {Braginski}(2006)}]{clarke2006squid}%
  \BibitemOpen
  \bibfield  {author} {\bibinfo {author} {\bibfnamefont {J.}~\bibnamefont
  {Clarke}}\ and\ \bibinfo {author} {\bibfnamefont {A.~I.}\ \bibnamefont
  {Braginski}},\ }\href@noop {} {\bibfield  {journal} {\bibinfo  {journal} {The
  SQUID handbook: Applications of SQUIDs and SQUID systems, John Wiley \&
  Sons}\ } (\bibinfo {year} {2006})}\BibitemShut {NoStop}%
\bibitem [{\citenamefont {Likharev}(1979)}]{likharev1979superconducting}%
  \BibitemOpen
  \bibfield  {author} {\bibinfo {author} {\bibfnamefont {K.}~\bibnamefont
  {Likharev}},\ }\href@noop {} {\bibfield  {journal} {\bibinfo  {journal}
  {Reviews of Modern Physics}\ }\textbf {\bibinfo {volume} {51}},\ \bibinfo
  {pages} {101} (\bibinfo {year} {1979})}\BibitemShut {NoStop}%
\bibitem [{\citenamefont {Tinkham}(1996)}]{tinkham1996introduction}%
  \BibitemOpen
  \bibfield  {author} {\bibinfo {author} {\bibfnamefont {M.}~\bibnamefont
  {Tinkham}},\ }\href@noop {} {\bibfield  {journal} {\bibinfo  {journal}
  {Introduction to superconductivity, Courier Corporation}\ } (\bibinfo {year}
  {1996})}\BibitemShut {NoStop}%
\bibitem [{\citenamefont {Podd}\ \emph {et~al.}(2007)\citenamefont {Podd},
  \citenamefont {Hutchinson}, \citenamefont {Williams},\ and\ \citenamefont
  {Hasko}}]{podd2007micro}%
  \BibitemOpen
  \bibfield  {author} {\bibinfo {author} {\bibfnamefont {G.}~\bibnamefont
  {Podd}}, \bibinfo {author} {\bibfnamefont {G.}~\bibnamefont {Hutchinson}},
  \bibinfo {author} {\bibfnamefont {D.}~\bibnamefont {Williams}}, \ and\
  \bibinfo {author} {\bibfnamefont {D.}~\bibnamefont {Hasko}},\ }\href@noop {}
  {\bibfield  {journal} {\bibinfo  {journal} {Physical Review B}\ }\textbf
  {\bibinfo {volume} {75}},\ \bibinfo {pages} {134501} (\bibinfo {year}
  {2007})}\BibitemShut {NoStop}%
\bibitem [{\citenamefont {Gumann}\ \emph {et~al.}(2007)\citenamefont {Gumann},
  \citenamefont {Dahm},\ and\ \citenamefont
  {Schopohl}}]{gumann2007microscopic}%
  \BibitemOpen
  \bibfield  {author} {\bibinfo {author} {\bibfnamefont {A.}~\bibnamefont
  {Gumann}}, \bibinfo {author} {\bibfnamefont {T.}~\bibnamefont {Dahm}}, \ and\
  \bibinfo {author} {\bibfnamefont {N.}~\bibnamefont {Schopohl}},\ }\href@noop
  {} {\bibfield  {journal} {\bibinfo  {journal} {Physical Review B}\ }\textbf
  {\bibinfo {volume} {76}},\ \bibinfo {pages} {064529} (\bibinfo {year}
  {2007})}\BibitemShut {NoStop}%
\bibitem [{\citenamefont {Hasselbach}\ \emph {et~al.}(2000)\citenamefont
  {Hasselbach}, \citenamefont {Veauvy},\ and\ \citenamefont
  {Mailly}}]{hasselbach2000microsquid}%
  \BibitemOpen
  \bibfield  {author} {\bibinfo {author} {\bibfnamefont {K.}~\bibnamefont
  {Hasselbach}}, \bibinfo {author} {\bibfnamefont {C.}~\bibnamefont {Veauvy}},
  \ and\ \bibinfo {author} {\bibfnamefont {D.}~\bibnamefont {Mailly}},\
  }\href@noop {} {\bibfield  {journal} {\bibinfo  {journal} {Physica C:
  Superconductivity}\ }\textbf {\bibinfo {volume} {332}},\ \bibinfo {pages}
  {140} (\bibinfo {year} {2000})}\BibitemShut {NoStop}%
\bibitem [{\citenamefont {Faucher}\ \emph {et~al.}(2002)\citenamefont
  {Faucher}, \citenamefont {Fournier}, \citenamefont {Pannetier}, \citenamefont
  {Thirion}, \citenamefont {Wernsdorfer}, \citenamefont {Villegier},\ and\
  \citenamefont {Bouchiat}}]{faucher2002niobium}%
  \BibitemOpen
  \bibfield  {author} {\bibinfo {author} {\bibfnamefont {M.}~\bibnamefont
  {Faucher}}, \bibinfo {author} {\bibfnamefont {T.}~\bibnamefont {Fournier}},
  \bibinfo {author} {\bibfnamefont {B.}~\bibnamefont {Pannetier}}, \bibinfo
  {author} {\bibfnamefont {C.}~\bibnamefont {Thirion}}, \bibinfo {author}
  {\bibfnamefont {W.}~\bibnamefont {Wernsdorfer}}, \bibinfo {author}
  {\bibfnamefont {J.}~\bibnamefont {Villegier}}, \ and\ \bibinfo {author}
  {\bibfnamefont {V.}~\bibnamefont {Bouchiat}},\ }\href@noop {} {\bibfield
  {journal} {\bibinfo  {journal} {Physica C: Superconductivity}\ }\textbf
  {\bibinfo {volume} {368}},\ \bibinfo {pages} {211} (\bibinfo {year}
  {2002})}\BibitemShut {NoStop}%
\bibitem [{\citenamefont {Hutchinson}\ \emph {et~al.}(2003)\citenamefont
  {Hutchinson}, \citenamefont {Qin}, \citenamefont {Kang}, \citenamefont {Lee},
  \citenamefont {Hasko}, \citenamefont {Blamire},\ and\ \citenamefont
  {Williams}}]{hutchinson2003hot}%
  \BibitemOpen
  \bibfield  {author} {\bibinfo {author} {\bibfnamefont {G.}~\bibnamefont
  {Hutchinson}}, \bibinfo {author} {\bibfnamefont {H.}~\bibnamefont {Qin}},
  \bibinfo {author} {\bibfnamefont {D.}~\bibnamefont {Kang}}, \bibinfo {author}
  {\bibfnamefont {S.}~\bibnamefont {Lee}}, \bibinfo {author} {\bibfnamefont
  {D.}~\bibnamefont {Hasko}}, \bibinfo {author} {\bibfnamefont
  {M.}~\bibnamefont {Blamire}}, \ and\ \bibinfo {author} {\bibfnamefont
  {D.}~\bibnamefont {Williams}},\ }\href@noop {} {\bibfield  {journal}
  {\bibinfo  {journal} {Superconductor Science and Technology}\ }\textbf
  {\bibinfo {volume} {16}},\ \bibinfo {pages} {1544} (\bibinfo {year}
  {2003})}\BibitemShut {NoStop}%
\bibitem [{\citenamefont {Hutchinson}\ \emph
  {et~al.}(2004{\natexlab{a}})\citenamefont {Hutchinson}, \citenamefont {Qin},
  \citenamefont {Hasko}, \citenamefont {Kang},\ and\ \citenamefont
  {Williams}}]{hutchinson2004controlled}%
  \BibitemOpen
  \bibfield  {author} {\bibinfo {author} {\bibfnamefont {G.}~\bibnamefont
  {Hutchinson}}, \bibinfo {author} {\bibfnamefont {H.}~\bibnamefont {Qin}},
  \bibinfo {author} {\bibfnamefont {D.}~\bibnamefont {Hasko}}, \bibinfo
  {author} {\bibfnamefont {D.}~\bibnamefont {Kang}}, \ and\ \bibinfo {author}
  {\bibfnamefont {D.}~\bibnamefont {Williams}},\ }\href@noop {} {\bibfield
  {journal} {\bibinfo  {journal} {Applied physics letters}\ }\textbf {\bibinfo
  {volume} {84}},\ \bibinfo {pages} {136} (\bibinfo {year}
  {2004}{\natexlab{a}})}\BibitemShut {NoStop}%
\bibitem [{\citenamefont {Hutchinson}\ \emph
  {et~al.}(2004{\natexlab{b}})\citenamefont {Hutchinson}, \citenamefont {Qin},
  \citenamefont {Kang}, \citenamefont {Lee}, \citenamefont {Blamire},
  \citenamefont {Hasko},\ and\ \citenamefont
  {Williams}}]{hutchinson2004fabrication}%
  \BibitemOpen
  \bibfield  {author} {\bibinfo {author} {\bibfnamefont {G.}~\bibnamefont
  {Hutchinson}}, \bibinfo {author} {\bibfnamefont {H.}~\bibnamefont {Qin}},
  \bibinfo {author} {\bibfnamefont {D.}~\bibnamefont {Kang}}, \bibinfo {author}
  {\bibfnamefont {S.}~\bibnamefont {Lee}}, \bibinfo {author} {\bibfnamefont
  {M.}~\bibnamefont {Blamire}}, \bibinfo {author} {\bibfnamefont
  {D.}~\bibnamefont {Hasko}}, \ and\ \bibinfo {author} {\bibfnamefont
  {D.}~\bibnamefont {Williams}},\ }\href@noop {} {\bibfield  {journal}
  {\bibinfo  {journal} {Microelectronic engineering}\ }\textbf {\bibinfo
  {volume} {73}},\ \bibinfo {pages} {773} (\bibinfo {year}
  {2004}{\natexlab{b}})}\BibitemShut {NoStop}%
\bibitem [{\citenamefont {Russo}\ \emph {et~al.}(2012)\citenamefont {Russo},
  \citenamefont {Granata}, \citenamefont {Esposito}, \citenamefont {Peddis},
  \citenamefont {Cannas},\ and\ \citenamefont
  {Vettoliere}}]{russo2012nanoparticle}%
  \BibitemOpen
  \bibfield  {author} {\bibinfo {author} {\bibfnamefont {R.}~\bibnamefont
  {Russo}}, \bibinfo {author} {\bibfnamefont {C.}~\bibnamefont {Granata}},
  \bibinfo {author} {\bibfnamefont {E.}~\bibnamefont {Esposito}}, \bibinfo
  {author} {\bibfnamefont {D.}~\bibnamefont {Peddis}}, \bibinfo {author}
  {\bibfnamefont {C.}~\bibnamefont {Cannas}}, \ and\ \bibinfo {author}
  {\bibfnamefont {A.}~\bibnamefont {Vettoliere}},\ }\href@noop {} {\bibfield
  {journal} {\bibinfo  {journal} {Applied Physics Letters}\ }\textbf {\bibinfo
  {volume} {101}},\ \bibinfo {pages} {122601} (\bibinfo {year}
  {2012})}\BibitemShut {NoStop}%
\bibitem [{\citenamefont {Hazra}\ \emph {et~al.}(2014)\citenamefont {Hazra},
  \citenamefont {Kirtley},\ and\ \citenamefont {Hasselbach}}]{hazra2014nano}%
  \BibitemOpen
  \bibfield  {author} {\bibinfo {author} {\bibfnamefont {D.}~\bibnamefont
  {Hazra}}, \bibinfo {author} {\bibfnamefont {J.~R.}\ \bibnamefont {Kirtley}},
  \ and\ \bibinfo {author} {\bibfnamefont {K.}~\bibnamefont {Hasselbach}},\
  }\href@noop {} {\bibfield  {journal} {\bibinfo  {journal} {Applied Physics
  Letters}\ }\textbf {\bibinfo {volume} {104}},\ \bibinfo {pages} {152603}
  (\bibinfo {year} {2014})}\BibitemShut {NoStop}%
\bibitem [{\citenamefont {Paul}\ \emph {et~al.}(2016)\citenamefont {Paul},
  \citenamefont {Biswas},\ and\ \citenamefont {Gupta}}]{paul2016micron}%
  \BibitemOpen
  \bibfield  {author} {\bibinfo {author} {\bibfnamefont {S.}~\bibnamefont
  {Paul}}, \bibinfo {author} {\bibfnamefont {S.}~\bibnamefont {Biswas}}, \ and\
  \bibinfo {author} {\bibfnamefont {A.~K.}\ \bibnamefont {Gupta}},\ }\href@noop
  {} {\bibfield  {journal} {\bibinfo  {journal} {Superconductor Science and
  Technology}\ }\textbf {\bibinfo {volume} {30}},\ \bibinfo {pages} {025017}
  (\bibinfo {year} {2016})}\BibitemShut {NoStop}%
\bibitem [{\citenamefont {McCaughan}\ \emph {et~al.}(2016)\citenamefont
  {McCaughan}, \citenamefont {Zhao},\ and\ \citenamefont
  {Berggren}}]{mccaughan2016nanosquid}%
  \BibitemOpen
  \bibfield  {author} {\bibinfo {author} {\bibfnamefont {A.~N.}\ \bibnamefont
  {McCaughan}}, \bibinfo {author} {\bibfnamefont {Q.}~\bibnamefont {Zhao}}, \
  and\ \bibinfo {author} {\bibfnamefont {K.~K.}\ \bibnamefont {Berggren}},\
  }\href@noop {} {\bibfield  {journal} {\bibinfo  {journal} {Scientific
  reports}\ }\textbf {\bibinfo {volume} {6}},\ \bibinfo {pages} {28095}
  (\bibinfo {year} {2016})}\BibitemShut {NoStop}%
\bibitem [{\citenamefont {Wu}\ \emph {et~al.}(2017)\citenamefont {Wu},
  \citenamefont {Chen}, \citenamefont {Wang}, \citenamefont {Wang},
  \citenamefont {Wo}, \citenamefont {Zhao}, \citenamefont {Liu}, \citenamefont
  {Wu},\ and\ \citenamefont {Wang}}]{wu2017measurement}%
  \BibitemOpen
  \bibfield  {author} {\bibinfo {author} {\bibfnamefont {L.}~\bibnamefont
  {Wu}}, \bibinfo {author} {\bibfnamefont {L.}~\bibnamefont {Chen}}, \bibinfo
  {author} {\bibfnamefont {H.}~\bibnamefont {Wang}}, \bibinfo {author}
  {\bibfnamefont {Q.}~\bibnamefont {Wang}}, \bibinfo {author} {\bibfnamefont
  {H.}~\bibnamefont {Wo}}, \bibinfo {author} {\bibfnamefont {J.}~\bibnamefont
  {Zhao}}, \bibinfo {author} {\bibfnamefont {X.}~\bibnamefont {Liu}}, \bibinfo
  {author} {\bibfnamefont {X.}~\bibnamefont {Wu}}, \ and\ \bibinfo {author}
  {\bibfnamefont {Z.}~\bibnamefont {Wang}},\ }\href@noop {} {\bibfield
  {journal} {\bibinfo  {journal} {Superconductor Science and Technology}\
  }\textbf {\bibinfo {volume} {30}},\ \bibinfo {pages} {074011} (\bibinfo
  {year} {2017})}\BibitemShut {NoStop}%
\bibitem [{\citenamefont {Biswas}\ \emph {et~al.}(2017)\citenamefont {Biswas},
  \citenamefont {Winkelmann}, \citenamefont {Courtois},\ and\ \citenamefont
  {Gupta}}]{biswas2017josephson}%
  \BibitemOpen
  \bibfield  {author} {\bibinfo {author} {\bibfnamefont {S.}~\bibnamefont
  {Biswas}}, \bibinfo {author} {\bibfnamefont {C.~B.}\ \bibnamefont
  {Winkelmann}}, \bibinfo {author} {\bibfnamefont {H.}~\bibnamefont
  {Courtois}}, \ and\ \bibinfo {author} {\bibfnamefont {A.~K.}\ \bibnamefont
  {Gupta}},\ }\href@noop {} {\bibfield  {journal} {\bibinfo  {journal} {arXiv
  preprint arXiv:1709.02569}\ } (\bibinfo {year} {2017})}\BibitemShut {NoStop}%
\bibitem [{\citenamefont {Russo}\ \emph {et~al.}(2016)\citenamefont {Russo},
  \citenamefont {Esposito}, \citenamefont {Crescitelli}, \citenamefont
  {Di~Gennaro}, \citenamefont {Granata}, \citenamefont {Vettoliere},
  \citenamefont {Cristiano},\ and\ \citenamefont
  {Lisitskiy}}]{russo2016nanosquids}%
  \BibitemOpen
  \bibfield  {author} {\bibinfo {author} {\bibfnamefont {R.}~\bibnamefont
  {Russo}}, \bibinfo {author} {\bibfnamefont {E.}~\bibnamefont {Esposito}},
  \bibinfo {author} {\bibfnamefont {A.}~\bibnamefont {Crescitelli}}, \bibinfo
  {author} {\bibfnamefont {E.}~\bibnamefont {Di~Gennaro}}, \bibinfo {author}
  {\bibfnamefont {C.}~\bibnamefont {Granata}}, \bibinfo {author} {\bibfnamefont
  {A.}~\bibnamefont {Vettoliere}}, \bibinfo {author} {\bibfnamefont
  {R.}~\bibnamefont {Cristiano}}, \ and\ \bibinfo {author} {\bibfnamefont
  {M.}~\bibnamefont {Lisitskiy}},\ }\href@noop {} {\bibfield  {journal}
  {\bibinfo  {journal} {Superconductor Science and Technology}\ }\textbf
  {\bibinfo {volume} {30}},\ \bibinfo {pages} {024009} (\bibinfo {year}
  {2016})}\BibitemShut {NoStop}%
\bibitem [{\citenamefont {Henkels}\ and\ \citenamefont
  {Kircher}(1977)}]{henkels1977penetration}%
  \BibitemOpen
  \bibfield  {author} {\bibinfo {author} {\bibfnamefont {W.}~\bibnamefont
  {Henkels}}\ and\ \bibinfo {author} {\bibfnamefont {C.}~\bibnamefont
  {Kircher}},\ }\href@noop {} {\bibfield  {journal} {\bibinfo  {journal} {IEEE
  Transactions on magnetics}\ }\textbf {\bibinfo {volume} {13}},\ \bibinfo
  {pages} {63} (\bibinfo {year} {1977})}\BibitemShut {NoStop}%
\bibitem [{\citenamefont {Barone}\ and\ \citenamefont
  {Paterno}(1982)}]{barone1982physics}%
  \BibitemOpen
  \bibfield  {author} {\bibinfo {author} {\bibfnamefont {A.}~\bibnamefont
  {Barone}}\ and\ \bibinfo {author} {\bibfnamefont {G.}~\bibnamefont
  {Paterno}},\ }\href@noop {} {\bibfield  {journal} {\bibinfo  {journal}
  {Physics and applications of the Josephson effect, Wiley Online Library}\
  }\textbf {\bibinfo {volume} {1}} (\bibinfo {year} {1982})}\BibitemShut
  {NoStop}%
\bibitem [{\citenamefont {Day}\ \emph {et~al.}(2003)\citenamefont {Day},
  \citenamefont {LeDuc}, \citenamefont {Mazin}, \citenamefont {Vayonakis},\
  and\ \citenamefont {Zmuidzinas}}]{day2003broadband}%
  \BibitemOpen
  \bibfield  {author} {\bibinfo {author} {\bibfnamefont {P.~K.}\ \bibnamefont
  {Day}}, \bibinfo {author} {\bibfnamefont {H.~G.}\ \bibnamefont {LeDuc}},
  \bibinfo {author} {\bibfnamefont {B.~A.}\ \bibnamefont {Mazin}}, \bibinfo
  {author} {\bibfnamefont {A.}~\bibnamefont {Vayonakis}}, \ and\ \bibinfo
  {author} {\bibfnamefont {J.}~\bibnamefont {Zmuidzinas}},\ }\href@noop {}
  {\bibfield  {journal} {\bibinfo  {journal} {Nature}\ }\textbf {\bibinfo
  {volume} {425}},\ \bibinfo {pages} {817} (\bibinfo {year}
  {2003})}\BibitemShut {NoStop}%
\bibitem [{\citenamefont {Annunziata}\ \emph {et~al.}(2010)\citenamefont
  {Annunziata}, \citenamefont {Santavicca}, \citenamefont {Frunzio},
  \citenamefont {Catelani}, \citenamefont {Rooks}, \citenamefont {Frydman},\
  and\ \citenamefont {Prober}}]{annunziata2010tunable}%
  \BibitemOpen
  \bibfield  {author} {\bibinfo {author} {\bibfnamefont {A.~J.}\ \bibnamefont
  {Annunziata}}, \bibinfo {author} {\bibfnamefont {D.~F.}\ \bibnamefont
  {Santavicca}}, \bibinfo {author} {\bibfnamefont {L.}~\bibnamefont {Frunzio}},
  \bibinfo {author} {\bibfnamefont {G.}~\bibnamefont {Catelani}}, \bibinfo
  {author} {\bibfnamefont {M.~J.}\ \bibnamefont {Rooks}}, \bibinfo {author}
  {\bibfnamefont {A.}~\bibnamefont {Frydman}}, \ and\ \bibinfo {author}
  {\bibfnamefont {D.~E.}\ \bibnamefont {Prober}},\ }\href@noop {} {\bibfield
  {journal} {\bibinfo  {journal} {Nanotechnology}\ }\textbf {\bibinfo {volume}
  {21}},\ \bibinfo {pages} {445202} (\bibinfo {year} {2010})}\BibitemShut
  {NoStop}%
\end{thebibliography}%

\end{document}